\documentclass[11pt]{article}
\usepackage{color}
\usepackage{cite}
\usepackage{epsfig}
\usepackage{amsfonts}
\usepackage{amssymb}

\definecolor{pink}{rgb}{1,.4,.7}
\definecolor{magenta}{rgb}{1,0,1}
\definecolor{violet}{rgb}{.9,.25,.6}
\definecolor{darkolivegreen3}{rgb}{.6,.8,.35}
\definecolor{maroon3}{rgb}{.8,.26,.56}
\definecolor{mediumorchid}{rgb}{.73,.33,.83}
\definecolor{mediumorchid1}{rgb}{1.,.33,.63}
\definecolor{darkgreen}{rgb}{0.1,.6,.13}
\definecolor{lightyellow}{rgb}{1.,1.,.82}
\definecolor{turquoise}{rgb}{.35,.80,.71}
\definecolor{coral}{rgb}{1.,.6,.21}
\definecolor{orangered}{rgb}{1.,.5,0.}
\definecolor{orange}{rgb}{1.,.65,.1}
\definecolor{blue1}{rgb}{.48,.53,1.}
\definecolor{gold}{rgb}{1.,.85,0.}
\definecolor{darkviolet}{rgb}{.54,.04,.84}

\def\ktwo{{\hat{\kappa}}^2}
\def\kfour{{\hat{\kappa}}^4}
\def\lambrs{{\Lambda}_{RS}}

\def\etal{{\it et al. }}
\def\rf#1{(\ref {#1})}
\def\Journal#1#2#3#4{{#1} {\bf #2}, (#3) #4}

\def\APP{\em Astropart. Phys.}

\def\CQG{\em Class.Quant.Grav.}

\def\IMA{{\em Int. J. Mod. Phys.} A}
\def\JHE{\em J. High Ener. Phys.}

\def\LNP{\em Lect. Notes Phys.}
\def\LRR{\em Living Rev.Rel.}

\def\NPB{{\em Nucl. Phys.} B}
\def\PLB{{\em Phys. Lett.}  B}
\def\PRL{\em Phys. Rev. Lett.}
\def\PRD{{\em Phys. Rev.} D}

\def\PUS{\em Phys. Usp.}

\def\ZPH{\em Z. Phys.}

\def\be{\begin{equation}}
\def\ee{\end{equation}}
\def\bea{\begin{eqnarray}}
\def\eea{\end{eqnarray}}
\begin{document}
\begin{center}
\Large \bf {Testing Brane World Models with Ultra High Energy Cosmic Rays}
\end{center}

\begin{center}
{\it Houri Ziaeepour\\
{Mullard Space Science Laboratory (MSSL)\\Holmbury St.Mary, Dorking RH5 6NT\\
Surrey, UK.\\
Email: {\tt hz@mssl.ucl.uk.ac}}
}
\end{center}

\begin {abstract}
The arrival time coherence of particles in the Ultra High Energy Air 
Showers where the center of mass energy of the interaction is of the order of 
$10^{15} eV$, puts strict constraint on the propagation of particles in a 
hypothetical extra-dimension. We first argue that at such high energies 
bulk modes and massive KK-modes can be produced abundantly and in many 
models their phase space 
volume is larger than confined modes. Then, we study the minimum 
propagation time in one and two-brane models and show that a large part of 
the parameter space of these models are ruled out unless the 
confinement of fields is proteced by symmetries up to energies 
not accessible even to the high energy tail of Ultar High Energy Cosmic Rays 
(UHECRs). 
As a by-product we confirm the result obtained in some previous works about 
the close relation between a small Cosmological Constant and the hierarchy 
problem.
\end{abstract}

%\keywords{Large Extra-Dimensions, Ultar High Energy Cosmic Rays}

\section {Introduction}
Ever since the proposition by Th. Kaluza and O. Klein in 1920s to use a 
5-dimensional space-time for unification of Gravity with Electromagnetism 
~\cite{kalklein}, space-times with more than $4$ dimensions have been the 
hope of physicists to solve problems of High Energy Particle Physics. 
Last ideas in 
these series are suggestions by N. Arkani-Hamed, \etal~\cite{nima} and by 
L. Randall and R. Sundrum~\cite{rs} for using large extra dimensions and 
localized matter to solve mass hierarchy problem inspired by some previous 
works of V. Rubakov and M. Shaposhinkov~\cite{domainwall} on domain walls in 
higher dimensional spaces and P. Horava and E. Witten~\cite{ads5z2} on 
M-theory models with Compactification in spaces with D-brane boundaries.

In the first proposals only gravity could propagate in the bulk. It has been 
however found that the total localization of all fields except graviton on 
3-branes is not realistic. In fact brane solution are cosmologically unstable 
and at least one scalar bulk field (radion)~\cite{radion}~\cite{instab} is 
necessary to stabilize the distance between branes. In some brane models 
inflaton~\cite{infbrane} also has to propagate to the bulk to make inflation 
with necessary properties. A deeper insight to the propagation of 
gravitational waves and massive particles with bulk modes in models with 
infinite bulk has illustrated that even the warping of the bulk can not stop 
their escape from branes~\cite{geod}~\cite{shortcut}.

Most of localization mechanisms are evolutionary i.e. based on special 
configuration of matter fields with localized properties like topological 
defect solutions which can arise during phase transitions in the Early 
Universe~\cite{defect}. In these models the real dimensionality  of the 
space-time is larger than $4$ but our Universe is confined to a domain wall 
(3-brane) where fields 
specially at low energies (with respect to quantum gravity scale) are 
geometrically or gravitationally localized. However, geometrical settings
like a warped metric are not always enough to confine fields on the branes. 
It has been shown that in spaces with $\geq 3$ extra-dimensions gravity can 
not be geometrically confined to a 3-brane and p-form fields in the bulk 
must be added to stabilize the brane (defect)~\cite{defect3}. In 5-dim. 
models gravity and scalar fields can be localized on the brane with negative 
tension~\cite{localize} (or the brane with smallest value of warp factor 
in 2-brane models of ~\cite{kantres}~\cite{houribr}). Warp geometry can 
localize fermions only on the positive tension brane (which can not solve 
the hierarchy problem). Vector fields can not 
be geometrically confined. Localization of gauge fields and fermions on the 
negative tension brane is achievable through special particle physics 
setups~\cite{localize}~\cite{rubakovvect}. The localization scale however is 
considered to be not much higher than warping scale, otherwise a new 
hierarchy can appear~\cite{vecopac}. At higher energies one expects that 
symmetry restoration (e.g. chiral symmetry of fermions) leads to escape of 
particles from the brane.

Even when a warped geometry is enough to confine fields on the brane, the 
wave function of the zero mode can penetrate to the bulk (but has an 
exponential maximum on the brane)~\cite{rs}. In infinite bulk models KK 
continuum 
begins from $m = 0$ and this affects the long range behavior of gravity and 
other massless fields~\cite{geod}~\cite{escape}~\cite{escape0}. Massive fields if they have 
bulk modes can decay to the bulk with a life time which depends on the 
fundamental scale of gravity~\cite{escape}~\cite{escape0}. For orbifoldized 
models the spectrum of KK modes is discrete. The long range effect of 
massive graviton modes is less important but the probability of decay of 
massive modes to the bulk is unchanged (see next section). 
Universal extra-dimension models in which all SM particles 
propagate to the compact dimensions are not ruled out for compactification 
radius of order $TeV^{-1}$ or even lower~\cite{universal}.

At present brane world models have been constrained only based on the 
probability of direct observation of processes involving the production of 
gravitons and its Kaluza-Klein 
modes~\cite {obsgrav0}~\cite {cosmoprob}~\cite {obsgrav2}. A detail 
investigation of observable signal of the RS type 
models in Tevatron and LHC are performed 
in~\cite {obsgrav3} and KK-mode production in the early Universe  
in~\cite {obsgrav4}. The existent and near future accelerators can 
constrain the scale of gravity (and thus the size of the extra-dimensions) 
up to $\sim 30 TeV$. The interaction of Ultra High Energy Cosmic Rays 
(UHECR) with protons in the terrestrial atmosphere has a CM energy close to 
$~ 1000 TeV$ and is the most energetic interaction of elementary particles 
we can study today. It can be used to constrain the fundamental scale of 
gravity and the compactification scale up to much higher energies.

The mass of KK modes detected by an observer on the brane is the result of 
smeared dimensions in the wave function. Classically however, it can be 
interpreted as a delay in the displacement of particles. For the observer 
on the brane if the delay slightly modifies 
the propagation of the particle in the detector, it is interpreted as a 
larger particle mass, otherwise it is seen as an arrival delay, specially 
with respect to the particles which propagate only on the brane. In the case 
of an air shower, the time coherence of the showers will be destroyed.

In this work we calculate the minimum propagation time of particles 
ejected to extra-dimensions for a number of warped brane models and compare 
them with arrival time resolution of present Air Shower detectors. We 
restrict our study to $D = 5$ models. This is enough for 
understanding the general characteristics of the propagation in 
extra-dimension from the point of view of an observer on a $(3+1)$-brane and 
can be considered as a special configuration for models with higher 
dimensions. Our attention is mostly concentrated on the classical 
structure of the brane models because results are independent of the 
detail of quantum field contents and origin of the branes. However, before 
doing this we must assess the possibility and the probability that UHECRs' 
interaction in the atmosphere can produce bulk modes i.e. escaping particles. 
Given the rarity of UHECRs, only models in 
which the probability of production of bulk modes is very high can be 
constrained by this method.

\section {Production of Bulk Modes} \label {secescape}
The whole idea of constraining brane models with UHECRs depends on the 
possibility and probability that remnants of UHECR interaction in the 
atmosphere can penetrate into the extra-dimensions. In this section we 
review the localization of fields on the brane with a special attention on 
models which provide bulk modes i.e. not all fields are confined to the 
visible brane at all energies. We estimate the probability of producing 
these modes at energy scale of interaction between UHECRs and nucleons in the 
terrestrial atmosphere.

By definition in universal models~\cite{universal} any field has bulk modes
and propagates in all space-time dimensions. The interesting case for 
solving the hierarchy problem is when the size of the compactified dimensions 
are of the order of weak interaction or lower\footnote{In some of universal 
models the 
extra-dimensions are not warped. Here we only study the propagation in 
warped spaces. Nevertheless, when the compactification scale $L^{-1}$ is 
much larger than ${\mu}$ (see Sec.~\rf{secmodels} for definition), the warp 
factor is very close to one and the results of following sections are 
applicable.}. This is $\sim 3$ 
orders of magnitude smaller than the CM energy of UHECRs' interaction. 
Therefore in the case of universal models, UHECRs can produce lowest 
KK-modes abundantly. In ~\cite {cosmoprob} 
it is argued that UHECRs can not probe the physics at very high energies 
simply because their interactions is dominantly electromagnetic. It is true 
that probability of the exchange of a heavy particle e.g a massive boson 
related to symmetries beyond Standard Model is very small with 
respect to a low transverse momentum EM cascade. However, in universal 
models as SM fields have bulk modes at very high energies all dimensions 
are ``seen'' as 
to be the same and the parameter space of EM cascades with non-zero momentum 
component in the extra-dimensions is much larger than cascades restricted to 
the three infinite dimensions. In the language of KK-modes, with a good 
accuracy the lowest modes can be considered as massless and they can be 
produced with the same probability as zero modes. In the following 
subsections we argue that for non-universal models one expects that at the 
CM energy of UHECRs interaction, most of localization mechanisms be no 
longer active and Standard Model particles can escape to the bulk.

\subsection {Scalars and Spin-2 Fields} \label {subsec:scalar}
In warped models scalars and spin-2 fields can be confined geometrically. The 
zero mode of these fields has an exponentially decreasing wave function in 
the bulk~\cite{rs}~\cite{greenfun}. For models with infinite bulk the 
KK-spectrum begins from zero and therefore there is not a real confined zero 
mode. In addition, if the field has a non-zero 5-dim. mass, it has been 
shown~\cite{shortcut} that 4-dim mass eigen modes on the brane are complex and 
decay to the bulk with a width:
\be
\Gamma / m_4 \propto (m_4/\mu)^2 \quad \quad m_4^2 = m_5^2 / 2 \label {mwidth}
\ee
where $m_4$ is the real part of the 4-dim. mass eigen value, $m_5$ is the 
5-dim. mass of the field, and $\mu$ is the warp scale 
in the static RS metric (Eq. \rf {rsmetric} below). If the fundamental scale 
of Quantum Gravity is comparable to the weak interaction scale, at CM energy 
of UHECRs interaction it is expected that due to radiative corrections even 
massless particles like gravitons have an effective non-zero mass. For short 
distances relevant to the propagation of UHECRs in the 
terrestrial atmosphere, massive modes have a Yukawa type potential and their 
coupling is exponentially suppressed~\cite{shortcut}. However, if the $m_4$ 
is much smaller than CM energy, the effect of exponential term in Yukawa 
potential is negligible. 
For most energetic UHECRs $E_{CM} \sim 10^{15} eV$. This means that the 
coupling to modes as massive as $1 TeV$ is roughly the same as massless 
modes. The width in \rf {mwidth} depends on the effective 5-dim. mass of the 
field and the warping scale. We discuss their implication on the decay of 
massive modes to the bulk and on the test of brane models in Section 
\ref {secmodels}.

The above argument is also true in the case of 2-brane models where the 
spectrum is discrete. We show briefly that in static/quasi-static models 
the zero 
mode of scalar/spin-2 fields with $m_5 > 0$ has an imaginary part i.e. it 
decays to the bulk (See also ~\cite {rubakovrev}). We determine the 
propagator of scalar/graviton using the 
Green function method discussed in detail in ~\cite{greenfun}. 

After changing variables $y$ in metric \rf {rsmetric} to:
\bea
z & \equiv & \frac{1}{\mu} e^{\mu y} \label {zdef} \\
ds^2 & = & \frac {R^2}{z^2} (\eta_{\mu \nu} dx^{\mu} dx^{\nu} - dz^2). 
\label {rsmetricz}
\eea 
we apply the boundary conditions to both branes. 
Without loss of generality we assume one of them is at $y = 0$ or 
$z = \frac{1}{\mu} \equiv R$ and the other at $y = L$ or 
$z = \frac{1}{\mu} e^{\mu L} \equiv R'$ \footnote {We are only interested 
in the case where bulk is static. Therefore implicitly it is assumed that 
these points correspond to the fixed points of the radion field.}. The Green 
function (2-point propagator) $\Delta (x, z, x', z')$ is the solution of 
4-dim. mass eigenstate equation:
\bea
(z^2 {\partial}_z^2 + z {\partial}_z + p^2 z^2 - d^2) 
\hat{\Delta}_p (z, z') & = & \frac {R^3}{z} \delta (z-z') \label {masseigen} \\
\Delta (x, z, x', z') & = & \int \frac {d^4 p}{(2 \pi) ^4} e^{ip (x - x')} 
{\Delta}_p (z, z') \label {fourierdelt} \\
{\Delta}_p (z, z') & \equiv & \biggl ( \frac {zz'}{R^2}\biggr )^2 
\hat{\Delta}_p (z, z') \label {deltahat} \\
d & = & \sqrt {4 + R^2 m_5^2} \label {ddef}
\eea
The boundary and matching conditions for right and left propagators:
\bea
\hat {\Delta}_< & \equiv & \hat{\Delta}_p (z, z') \quad z < z' \quad , \quad 
\hat {\Delta}_> \equiv \hat{\Delta}_p (z, z') \quad z > z' 
\label {leftright}
\eea
are as followings (boundary conditions are deduced from \rf {leftright} 
and matching condition from \rf {masseigen}):
\bea
{\partial}_z (z^2 \hat {\Delta}_>)|_{z = R'} & = & 0 \label {bondrp} \\
{\partial}_z (z^2 \hat {\Delta}_<)|_{z = R} & = & 0 \label {bondr} \\
\hat {\Delta}_<|_{z = z'} & = & \hat {\Delta}_>|_{z = z'} 
\label {matchcond0} \\
{\partial}_z (\hat {\Delta}_< - \hat {\Delta}_>)|_{z = z'} & = & 
\frac {R^3}{z'^3} \label {matchcond1}
\eea
Solutions of \rf {masseigen} are linear combination of Bessel Functions:
\be
A (z', R, R') J_{d} (pz) + B (z', R, R') N_{d} (pz) 
\label {gensol}
\ee
Applying the conditions \rf {bondrp}-\rf {matchcond1} to this 
solution leads to an equation which determines the KK mass spectrum:
\be
\frac {pR' J_{\nu} (pR') + (1 - \nu) J_{\nu + 1} (pR')}{pR' N_{\nu} (pR') + 
(1 - \nu) N_{\nu + 1} (pR')} = \frac {pR J_{\nu} (pR) + (1 - \nu) 
J_{\nu + 1} (pR)}{pR N_{\nu} (pR) + (1 - \nu) N_{\nu + 1} (pR)} \label {kksol}
\ee
with $\nu \equiv d - 1$. Finding the exact solution of \rf {kksol} is not 
trivial. To consider only a simple case we assume that 4-dim. mass of the 
scalar field 
$|p| \ll \mu = \frac {1}{R}$, i.e. $pR \ll 1$. Regarding the Standard Model, 
this can be applied to a confined Higgs when the scale of compactification 
is much higher than Higgs mass or to a light axion like scalar or to 
graviton with a small mass due to radiative corrections. 
We keep only lowest powers of $pR$ in the expansion of $J_{\nu}$. For solving
hierarchy problem $R'\gg R$. Using the asymptotic expansion of Bessel 
functions, \rf {kksol} reduces to:
\bea
{\eta}'{\eta}^{-(\nu + 1)} \sqrt {\frac {2}{\pi {\eta}'}} \biggl (\cos 
({\eta}' - \frac {\pi\nu}{2} - \frac {\pi}{4}) - \frac {1 - \nu}{{\eta}'} 
sin ({\eta}' - 
\frac {\pi\nu}{2} - \frac {\pi}{4}) \biggr ) && \nonumber \\
 \frac {{\eta}^2 (1 + \frac {\eta}{2 (\nu - 1)}) + (\nu -1)(2 \nu + \eta)}
{\nu 2^{-\nu}\sin (\nu\pi) \Gamma (-\nu)} & = & 0 \label {kkfinal}
\eea
where $\eta \equiv pR$ and ${\eta}' \equiv pR'$. For $\nu \gtrsim 1$ the 
solutions of \rf {kkfinal} lead to the following mass spectrum:
\bea
i p_0 & = & m_4 - i \Gamma \\
m_4 & \approx & \mu \sqrt {2 \nu (\nu - 1)} \approx \frac {m_5}{\sqrt {2}} 
\quad , \quad \Gamma \approx {m_5}^2 / 8 \mu \label {zeromode} \\
|p'_n| & \approx & \mu e^{-\mu L} (n \pi + \frac {\pi\nu}{2} + 
\frac {3\pi}{4}) \quad \quad \mbox {For large $n$.} \label {kkmode}
\eea
Continuity properties of $J_{\nu}$ guarantees that \rf {kkfinal} is also 
valid when $m_5 \rightarrow 0$ or equivalently $\nu \rightarrow 1$. In this 
case $\eta = 0$ or $p_0 = 0$.  For $pR \ll 1$ the mass difference between 
KK-modes $p_n$ is $\Delta p \propto 1/R' \ll 1/R = \mu$. Due to special 
properties of Bessel Function with integer index, the zero mode of massless 
fields is protected from decay even when the spectrum of KK-modes for massive 
particles begins roughly from zero. Tunneling probability depends on 5-dim. 
mass of the field and on warping scale $\mu$. If $\mu$ is large, the 
probability of zero-mode decay to the bulk can be small. However, except for 
very light particles a $\mu$ as large as $1 TeV$ provides an enough large 
width ($\Gamma > 10^{-10} eV$) for decay to the bulk during propagation 
in the terrestrial atmosphere if $m_5$ is in the mass range of SM particles.

To see the effect of mass on the coupling we can investigate the mass 
dependence of the propagator on the branes. Using \rf {leftright} and 
\rf {bondrp}-\rf {matchcond1}, one can determine the integration 
coefficients $A (z', R, R')$ and $B (z', R, R')$ in \rf {gensol} and right 
and left propagators:
\bea
\hat {\Delta}_< (z,z') & = & \frac {\pi R^3 \biggl({\Delta}_0 J_{\nu + 1} 
(pz') - {\Delta}_1 N_{\nu + 1} (pz') \biggr) \biggl ({\Delta}_2 
J_{\nu + 1} (pz) - 
{\Delta}_3 N_{\nu + 1} (pz) \biggr )}{z'^2 ({\Delta}_0 {\Delta}_3 - 
{\Delta}_1 {\Delta}_2)} \label {propleft}\\
\hat {\Delta}_> (z,z') & = & \frac {\pi R^3 \biggl({\Delta}_2 J_{\nu + 1} 
(pz') - {\Delta}_3 N_{\nu + 1} (pz')\biggr ) \biggl ({\Delta}_0 J_{\nu + 1} (pz) - 
{\Delta}_1 N_{\nu + 1} (pz)) \biggr )}{z'^2 ({\Delta}_0 {\Delta}_3 - 
{\Delta}_1 {\Delta}_2)}\label {propright}\\
{\Delta}_0 & \equiv & pR' N_{\nu} (pR') + (1 - \nu) N_{\nu + 1} (pR') \\
{\Delta}_1 & \equiv & pR' J_{\nu} (pR') + (1 - \nu) J_{\nu + 1} (pR') \\
{\Delta}_2 & \equiv & pR N_{\nu} (pR) + (1 - \nu) N_{\nu + 1} (pR) \\
{\Delta}_3 & \equiv & pR J_{\nu} (pR) + (1 - \nu) J_{\nu + 1} (pR)
\eea
Restricting these equations to the branes gives the 2-point propagators:
\bea
{\Delta}(x,R,x',R) & = & \frac {1}{(2\pi)^4} \int dp^4 e^{ip (x-x')} 
\frac {p^{-1}({\Delta}_0 J_{\nu + 1} (pR) - {\Delta}_1 N_{\nu + 1} (pR)) }
{({\Delta}_0 {\Delta}_3 - {\Delta}_1 {\Delta}_2)} \label {propr}\\
{\Delta}(x,R',x',R') & = & \frac {R'^2}{R^2 (2\pi)^4} \int dp^4 e^{ip (x-x')} 
\frac {p^{-1}({\Delta}_2 J_{\nu + 1} (pR') - {\Delta}_3 N_{\nu + 1} (pR')) }
{({\Delta}_0 {\Delta}_3 - {\Delta}_1 {\Delta}_2)} \label {proprp}
\eea
The term $R'^2/R^2$ in \rf {proprp} reflects the difference between metric on 
the branes. Consistently, the roots of dominator in \rf {propr} and 
\rf {proprp}, are the same as \rf {kksol} and correspond to KK-modes. 
Near each mass mode the propagators can be written as a Yukawa propagator 
with complex mass and a coupling equal to the residue of the integrand. 
Applying this procedure to \rf {proprp} one finds that on the brane at $R'$ 
for $\nu > 1$:
\be
\frac {g_0^2}{g_n^2} \sim \frac {\biggl |\frac {p_0}{\mu} \biggr |^{2\nu}}
{\biggl |\frac {{p'}_n}{\mu} \biggr |^{2(\nu - 1)}} \label {couple}
\ee
In our approximation $|p_0/\mu| \propto m_5 /\mu \ll 1$, 
$|p'_n/\mu| > |p_0/\mu|$ and it can be even larger than 1. Therefore 
coupling to KK-modes is larger than to zero-mode. The probability of 
production of zero-mode with respect to $n^{th}$ KK-mode is:
\be
V \sim \biggl |\frac {p_0}{{p'}_n}\biggr |^{2(\nu -1)} \biggl |\frac {{p'}_n}
{\mu} \biggr |^2 \label {probcouple}
\ee
For KK-modes with $|p'_n| \sim \mu$ the branching ratio $V < 1$ and therefore 
the probability of production of these modes is larger than zero-mode.

In conclusion, radiative corrections that can induce an effective mass for 
scalar/spin-2 fields weakens their confinement on the brane. In warped 
2-brane models even when $m_5 = 0$ the mass difference between the zero 
mode and massive KK-modes on the brane with smallest warp factor is small 
and they can be abundantly produced in high energy interactions.

\subsection {Fermions and Gauge Fields}
Fermions can not be localized to the $TeV$ scale brane (i.e. brane at $R'$) 
gravitationally but a 
chiral symmetry breaking can confine them~\cite {localize}. Their escape to 
the bulk when their $m_5 > 0$ has been studied in detail for one brane 
models in ~\cite {shortcut} and ~\cite {massfermion}. The detail of the 
formalism is very similar to the case of a scalar field except that the mass 
eigenstate equation includes a chiral mass term due to interaction of fermion modes with a 
bulk scalar responsible for the symmetry breaking. The width of the zero mode 
depends on the coupling between fermions and the inferred scalar field and 
without detail knowledge of underlying particle physics it is difficult to 
assess the probability of decay to the bulk. For 
2-brane models the situation should not be very different and general 
conclusions of Sec.\ref {subsec:scalar} must be applicable.

At present the particle physics models don't fix the scale of the symmetry 
breaking. For not creating a new hierarchy however it can not be much larger 
than compactification scale~\cite {vecopac} or fundamental scale of gravity 
$M_5$. Therefore, at energy scale of 
UHECRs interaction in the atmosphere, not only it is possible to produce 
KK-modes, it is very probable that at such energies the restoration of 
chiral symmetry completely removes the confinement of fermions and open 
the extra-dimension even to fermionic zero modes (presumably SM matter).

As for gauge bosons, the most successful scenario for their confinement on 
the branes is 
based on adding an induced kinetic term to the action of bulk gauge fields on 
the brane. It appears due to the interaction of these fields with confined 
charged scalar or fermions on 
the branes~\cite {rubakovvect}. Other suggestions are mostly 
equivalent to this scenario~\cite {gaugeequiv}. Once the charged fields 
become able to escape to the bulk, they drag their interaction vertex with 
gauge fields to the bulk and release them from confinement. It has been 
shown~\cite {ftrs} that the coupling of gauge boson KK-modes to fermions on 
the brane is stronger than coupling of their corresponding zero-mode 
(similar to the self coupling of scalars \rf {couple}). 

The general conclusion of this section is that regarding:{\bf\it
\begin {description}
\item {- } Very high CM energy of interaction of most energetic Cosmic 
Rays in the atmosphere which is much higher than natural confinement scale 
of Standard Model particles on the brane and natural fundamental scale of 
gravity to solve the hierarchy problem;
\item {- } The fact that confinement of SM fields is not intrinsic but the 
result of either a broken symmetry (for fermions) or interactions 
(for bosons);
\item {- } That the particle physics in 5-dim can not be completely massless 
and at least part of the particle spectrum must acquire mass as it is the 
case in observable 4-dim Universe. In fact as radion is in fact the scalar 
component of 5-dim. metric perturbations~\cite {radionpert}, it couples to 
all bulk fields and radiatively induces a small mass term. Consequently, 
zero-modes (presumably SM particles) are not stable and in a finite time 
decay to the bulk unless another phenomenon like symmetry breaking prevent it;
\end {description}
the phase space of production of bulk modes at high energy tail of UHECRs 
spectra seems to be higher than confined modes.} 

Until now more 
than 100 coherent air showers have been observed with $E_{CM} \gtrsim 300 TeV$
(assuming interaction with nucleons in the atmosphere), 17 between them have 
energies more than $450 TeV$ and one has a CM energy close to 
$1000 TeV$~\cite{wimpzilla}. Assuming that UHECRs interaction with 
$E_{CM} \gtrsim 300 TeV$ produces bulk modes abundantly, the mere 
observation of coherent showers up to energies close to $10^{21} eV$ 
constraints the parameter space of brane models.

As the assessment of cross-sections and other details are model dependent,  
in the rest of this work we simplify the problem of constraining brane 
models and consider only the classical propagation of 
the bulk modes. This method has been already used by other authors to study 
some of cosmological consequences of brane models~\cite{freez}~\cite{geod}.

\section {Propagation} \label {secprop}
The geodesic path of particles in the bulk has been already studied in a 
number of previous works~\cite{freez}~\cite{geod}~\cite{shortcut}
~\cite{horiz}. However, most of them are concerned with the 
possible acausality of paths for observer on a brane and their purpose is to 
see if it can solve the cosmological horizon problem in the early universe. 
Here we are concerned with the present 
evolution of the Universe and simplifying assumptions will be based on  
its present very slowly changing state.\\
The metric of the 5-dim brane models can be written as the following:
\be
ds^2 = n^2 (t, y) dt^2 - a^2 (t, y) \delta_{ij} dx^i dx^j - b^2 (t, y) dy^2. 
\label {metric}
\ee
For a static bulk $b^2 (t, y)$ is constant and we can normalize coordinates 
such that $b = 1$. The geodesic of a particle is defined by:
\bea
\frac {du^0}{d\tau} + \frac {\dot {n}}{n} u^0 u^0 + \frac {2n'}{n} u^4 u^0 + 
\frac {a\dot {a}}{n^2} u^i u^j {\delta}_{ij} & = & 0 \label {u0eq}\\
\frac {du^i}{d\tau} + \frac {2a'}{a} u^i u^4 + \frac {2\dot {a}}{a} u^i u^0 
 & = & 0 \label {uieq}\\
\frac {du^4}{d\tau} + nn' u^0 u^0 - aa' u^i u^j {\delta}_{ij} & = & 0 
\label{u4eq}
\eea
\be
u^0 = \frac {dt}{d\tau}, \quad\quad u^i = \frac {dx^i}{d\tau}, \quad\quad 
u^4 = \frac {dy}{d\tau}. \label {udef}
\ee
$\tau$ is the proper time parameter along the particle world line. It is 
easy to see that \rf{uieq} is integrable and:
\be
u^i = \frac {{\theta}^i}{a^2} \label {uisol}
\ee
where ${\theta}^i$ is an integration constant. As we are only interested in 
the minimum delay in the arrival of particles due to the propagation in the 
extra-dimensions, we put ${\theta}^i = 0$. Later we try to estimate 
qualitatively the effect of a non-zero ${\theta}^i$. 
Even after this simplification the system of equations \rf{u0eq}-\rf{u4eq} 
is highly non-linear and coupled. In the following we calculate an 
analytical solution 
for the case $\dot {n}/n \thickapprox 0$. This approximation is justified 
when we are interested in the propagation of particles in an extremely short 
period of time with respect to the expansion rate of the bulk or the brane. 
In fact from the solution of the Einstein equations~\cite{binetruy99}, 
$n (t, y)$ can be normalized such that:
\be
n (t, y) = \frac {\dot {a} (t,y)}{\dot {a}_0 (t)} \label {ndef}
\ee
where ${a}_0 (t) = a (t, y=0)$ assuming that one of the branes is at 
$y = 0$. At present, both 
$\dot {a} (t,y)$ and $\dot {a}_0(t)$ are very slowly varying quantities. 
Therefore $n (t, y) \sim n (t_0, y)$ where $t_0$ is the present time, and 
$\dot {n} (t, y) \sim 0$.\\
Under this approximation:
\be
\frac {dn}{d\tau} \thickapprox n'u^4
\ee
and:
\bea
\frac {du^0}{d\tau} + \frac {2u^0}{n}\frac {dn}{d\tau} = 0 & & \label 
{u0eqpa}\\
\frac {du^4}{d\tau} + \frac {n (u^0)^2}{u^4}\frac {dn}{d\tau} = 0 & & 
\label{u4eqpa}
\eea
The solution of \rf{u0eqpa} and \rf{u4eqpa} is straightforward:
\be
u^0 = \theta n^{-2}, \quad\quad\quad u^4 = \pm \biggl (\eta - \frac {{\theta}^2}{n^2} \biggr)^ {\frac {1}{2}} \label {solu}
\ee
The parameters $\theta$ and $\eta$ are integration constants and must be 
determined from the initial conditions. The $\pm$ sign defines the direction of the propagation. In the rest of this letter we neglect the direction and consider only the absolute value of $u^4$.\\
For a particle leaving the visible brane placed at $y = y_b$ at time 
$t = t_0$:
\be
\theta = u^0 (t_0,y_b) n^2 (t_0,y_b), \quad\quad\quad 
\eta = \frac {{\theta}^2}{n^2 (t_0,y_b)} - (u^4(t_0,y_b))^2 = 
\biggl \{^{\mbox{$1 \quad\quad$ Massive 
particles,}}_{\mbox{0 \quad\quad Massless particles}.} \label {ab}
\ee
After eliminating the proper time from $u^0$ and $u^4$, one obtains the 
equation of motion in the bulk (For simplicity we assume that $Dy/d\tau \approx dy/d\tau$ and $Dt/d\tau \approx dt/d\tau$):
\bea
\frac {dy}{dt} & = & \frac {n^2 (t, y)}{\theta} \sqrt {\frac {{\theta}^2}
{n^2 (t, y)} - \varepsilon} \label {eqm} \\
\varepsilon & \equiv & \biggl \{^{\mbox{$1 \quad\quad$ Massive 
particles,}}_{\mbox{0 \quad \quad Massless particles}.} \label {dydt}
\eea
Our approximations are valid only when $dy/dt$ is real. This put limits on 
the testable part of the parameter space of the models (see below).\\ 
Einstein equations give the solution for $a (t,y)$ and 
$n (t,y)$~\cite{binetruy99}. For a flat visible brane:
\bea
a^2 (t, y) & = & {\mathcal A}(t)\cosh (\mu y) + 
{\mathcal B}(t)\sinh (\mu y) + {\mathcal C}(t) \label {aa}, \\
\dot {a}^2 (t, y) & = & n^2 (t, y) a_0^2 (t) = \frac {\biggl (\dot{\mathcal A}(t)\cosh (\mu y) + 
\dot{\mathcal B}(t)\sinh (\mu y) + \dot{\mathcal C}(t)\biggr )^2}{4 a^2 (t, y)} 
\label {dotaa}, \\
{\mathcal A}(t) & = & {a_0}^2 (t) - {\mathcal C(t)} \label {aaa},\\
{\mathcal B}(t) & = & -{\rho}'_{b_0} {a_0}^2 (t), \label {bbb}\\
{\mathcal C}(t) & = & -\frac {2 {{\dot {a}}_0}^2 (t)}{{\mu}^2}, \label {ccc}\\
\mu & \equiv & \sqrt {\frac {2 \ktwo}{3} |{\rho}_B|}
\eea
For any density $\rho$, ${\rho}' \equiv \rho / \lambrs$, 
$\lambrs \equiv 3 \mu / \ktwo$. The densities ${\rho}'_{b_0}$ and ${\rho}_B$
are effective total energy density of the brane at $y=0$ and 
the bulk respectively. We consider only AdS bulk models with 
${\rho}_B < 0$. The constant $\ktwo = 8 \pi / M_5^3$ is the gravitational 
coupling in the 5-dim. space-time. The model dependent details like how 
${\rho}'_{b_0}$ and ${\rho}_B$ are related to the field contents in the 
bulk and on the brane and how they evolve are irrelevant for us as long as 
we assume a quasi-static model. 
The solution \rf{aa} is valid both for one brane and multi-brane 
models. The only difference between them is in the application of Israel 
junction conditions~\cite{kantres}~\cite{houribr}.\\
Equation \rf{dydt} is non-linear and its integration non-trivial. We use 
again the quasi-static properties of the present Universe and its low energy 
density to simplify the integration. Matter density on the branes at late 
time is much smaller than the brane tension or induced tension by scalar 
fields~\cite{brax}.
Therefore, it is not unreasonable to neglect time dependence of densities 
in \rf{aaa}-\rf{ccc} and 
to consider only cosmological constant type energy-momentum densities. 
This simplification is even more justified in our case where we have to 
deal only with very short duration of the propagation in the 
extra-dimensions. This approximation and \rf{aaa}-\rf{ccc} lead to:
\bea
\dot {\mathcal C} & = & -\frac {4 {\dot {a_0}}^3 (t)}{{\mu}^2 a_0 (t)} 
\label {dccc},\\
\dot {\mathcal A} & = & 2 a_0 (t){\dot {a}}_0 (t) - \dot {\mathcal C}(t), 
\label {daaa}\\
\dot {\mathcal B} & = & -2 {\rho}'_{b_0} a_0 (t){\dot {a}}_0 (t). \label {dbbb}
\eea
After changing variable $y$ to $z = e^{\mu y}$ and using \rf{ndef}:
\bea
\dot {a}^2 (t, z) & = & -\frac {{\mu}^2 {\mathcal C}(t)}{2z} {\mathcal D}
(t, z), \label {addef}\\
{\mathcal D} & \equiv & \frac {1}{2} \biggl [(1-{\rho}'_{b_0} - 
\frac {{\mathcal C}(t)}{a^2_0}) z^2 + \frac {2 {\mathcal C}(t)}{a^2_0}z + 
(1 + {\rho}'_{b_0} - \frac {{\mathcal C}(t)}{a^2_0}) \biggr ], 
\label {ddef}\\
\frac {dz}{dt} & = & \mu \sqrt {{\mathcal D} (z - \frac {\epsilon}
{{\theta}^2}{\mathcal D})}. \label {eqz}
\eea
If an ejected particle to the bulk comes back to the brane, $u^4$ must go to 
zero at some point in the bulk before the particle arrives to the bulk 
horizon (if it is present). The roots of \rf{eqz} correspond to these 
turning points and determine the propagation time in the bulk. 
In the next simplifying step we use again the fact that the typical 
propagation 
time we are interested in is very much shorter than the age of the Universe 
and therefore ${\mathcal A}, {\mathcal B}, {\mathcal C}$ and 
$\dot {\mathcal A}, \dot {\mathcal B}, \dot {\mathcal C}$ during propagation 
are roughly constant, the right hand side of \rf{eqz} depends 
only on $z$ and is easily integrable:
\be
{\Delta} t_{propag} \equiv 2 (t_{stop} - t_0) = \int_{z_0}^{z_{stop}} 
\frac {2 dz}{\mu \sqrt {{\mathcal D} (z - \frac {\epsilon}
{{\theta}^2}{\mathcal D})}} \label {tstop}
\ee
In \rf{tstop}, $t_0$ is the initial time of propagation in the 
extra-dimension and $t_{stop}$ is the time when the particle's velocity 
changes its direction, i.e. when $dz/dt = 0$. The integral in \rf{tstop} 
is related to the elliptical integrals of the first type 
${\mathcal F} (\omega, \nu)$ where $\omega$ and $\nu$ are analytical 
functions of the denominator roots in \rf{tstop} and 
$z_0$~\cite {integhb}. Note that $z_{stop}$ corresponds to the closest root 
to $z_0$.

\section {Test of Brane Models} \label {secmodels}
In this section we apply the formalism discussed in the previous section to 
most popular brane models and determine the propagation time of high 
energy particles in the fifth dimension. Note that the calculation of 
propagation time in these models under our approximations is valid only for 
durations very smaller than the age of the Universe and if in the following 
figures in part of the parameter space the propagation time can be larger, 
this part of the figure should not be considered.

\subsection {2-Brane Models}
\begin{figure}[t]
\begin{center}
\psfig{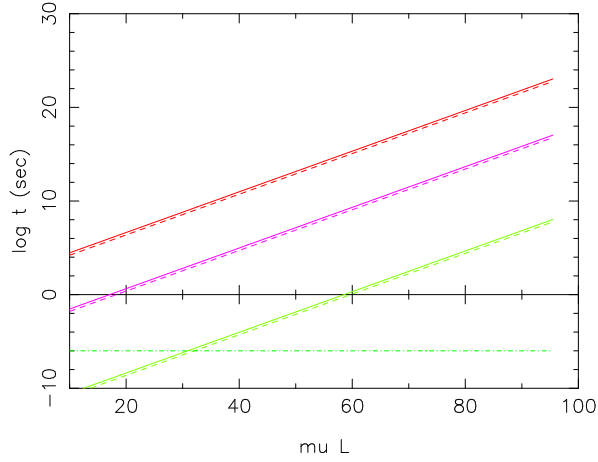}
\caption{Propagation time for relativistic particles with 
$u^0_L (t_0)/N = 10^3$ (full line) and $u^0_L (t_0)/N = 1.2$ (dash 
line) in RS model. Red, magenta and light green curves correspond to 
$M_5 = 10^{13} eV$, $M_5 = 10^{15} eV$ and $M_5 = 10^{18} eV$ (or $\mu \sim 
10^{-17} eV$, $\mu \sim 10^{-11} eV$ and $\mu \sim 10^{-2} eV$ for 
fine-tuned model) respectively. The dark green 
line shows the time coherence precision of present Air Shower detectors.
\label {fig:rstime}}
\end{center}
\end{figure}
It has been shown in ~\cite{kantres}~\cite{houribr} that by imposing 
constraints on the 
visible brane to obtain the observed value of cosmological constant 
$\Lambda$ and Newton coupling constant $G$ and to solve the hierarchy 
problem, all parameters of this class of models i.e. ${\rho}_0$, ${\rho}_L$ 
and $\lambrs$, can be determined as a function of $\mu L$ where $L$ is the 
distance 
between two branes. It is not however possible to find an exact analytical 
form for the solutions. Moreover, the analytical solution in 
~\cite{houribr} has been obtained 
for a special setup which decouples hidden and visible branes. 
Here we free some of the constraints, first to be able to find analytical 
solutions, and second to extend this study to a larger number of models.

\subsubsection {Geodesics in Static RS Models}
In the original RS model with static metric:
\be
ds^2 = e^{-\mu y}\eta_{\mu \nu} dx^{\mu} dx^{\nu} - dy^2. 
\label {rsmetric}
\ee
the main constraint on the model is the cancellation of 
the cosmological constant on the visible brane which leads to the equal and 
opposite sign tensions ${\rho}_0 = - {\rho}_L = \lambrs$. The solution of 
the hierarchy problem limits the range of the parameter $\mu L$. It has been 
shown~\cite{geod} that in the fine-tuned RS model photons leave the brane 
and never return. Assuming the visibility of the all dimensions of the 
space-time at very 
high energies, in this model we could never observe the ultra high energy 
particles and therefore it is automatically ruled out. Nonetheless, to test 
the formalism of the previous section we apply it to this model.\\
For a very small cosmological constant (as it is assumed in the 
process of fine-tuning) ${\mathcal C} (t_0) \approx 0$ and 
${\mathcal D} (t_0, z) \approx 1$. For massive particles, the denominator in 
\rf{tstop} has only one root: $z_1 = 1 /\theta$ and:
\be
{\Delta} t_{propag} = \frac {2 e^{\frac {\mu L}{2}}}{\mu} \sqrt {1 - 
\frac {e^{\mu L}}{(u^0_L (t_0))^2}} \quad \quad \quad \theta = e^{-\mu L} 
u^0_L (t_0). \label {deltrs}
\ee
Fig.\ref{fig:rstime} shows ${\Delta} t_{propag}$ as a function of $\mu L$ 
and $M5$ for massive relativistic particles. With present air shower 
detectors time resolution of order $10^{-6} sec$, only when $M_5 \gtrsim 
10^{18} eV$, the model is compatible with the observed time coherence of the 
UHE showers. For fine-tuned RS model $\mu \approx G /\ktwo$ ~\cite{rs} i.e. 
${\mu} = G M_5^3 \sim 10^{-3} eV$ for $M_5 \sim 10^{18} eV$ ~\cite{rs}. 
Due to smallness of $\mu$ and consequently lightness of KK-modes for SM 
particles even this model with large $M_5$ has already been ruled 
out~\cite {ftrs} unless a conserved quantum number prevents the production 
of KK-modes~\cite{universal}. 

For massless particles:
\be
z (t) - z_0 = \mu (t - t_0)^2 \label {zrs}
\ee
In \rf{zrs}, $z (t)$ is monotonically increasing and there is no stopping 
point. With our approximations there is no horizon in the bulk because 
$a (t, y)$ is roughly constant. Therefore \rf{zrs} means that massless 
particles simply continue their path to the hidden brane and their faith 
depends on what happen to them there. At very high CM energy of UHECR 
interactions 
if charged particles can escape to the bulk photons are also dragged to the 
bulk and never come back.

\subsubsection {General Solution}
Numerical solution of constrained 2-brane models in 
~\cite{kantres}~\cite{houribr} shows that for $\mu L \gtrsim 5$ the tension 
on both branes is positive and very close to $\lambrs$. We can use constraints 
on the Cosmological Constant and hierarchy to find ${\rho}'_0$ and 
${\rho}'_L = {\rho}'_b$. We redefine them as ${\rho}'_0 = 1 + \Delta 
{\rho}'_0$ and ${\rho}'_L = 1 + \Delta {\rho}'_L$. To solve hierarchy problem 
(See equations 29-31 in ~\cite{houribr}): 
\be
\frac {M^2_5}{M^2_{pl}} \sim N^2 \equiv \frac {n^2_L}{n^2_0} = 
\frac {{\rho}'_{{\Lambda}_0} (1-\cosh (\mu L)) + \sinh (\mu L)}{{\rho}'_{
{\Lambda}_L} (1-\cosh (\mu L)) + \sinh (\mu L)} \ll 1 \label {n2} 
\ee
This leads to:
\be
\Delta {\rho}'_0 = \frac {N^2 \biggl (1 - e^{-\mu L} + \Delta {\rho}'_L (1 - 
\cosh (\mu L) \biggr )}{1 - \cosh (\mu L)} - \frac {1 - e^{-\mu L}}{1 - 
\cosh (\mu L)} \label {deltrho0}
\ee
For a very small $N^2$ and $\Delta {\rho}'_L \lesssim 1$, the first term in 
\rf{deltrho0} is ${\mathcal O} (N^2)$ and:
\be 
\Delta {\rho}'_0 \approx - \frac {1 - e^{-\mu L}}{1 - \cosh (\mu L)} \approx 
- \frac {1}{1 - \cosh (\mu L)} \approx 2e^{-\mu L} \label {deltrho0p}
\ee
Using $\dot {a}^2_L / {a}^2_L = H^2$ where $H$ is the Hubble Constant on the 
visible brane~\cite{houribr}:
\bea
\Delta {\rho}'_L & = & \frac {1}{2 N^2 \sinh (\mu L)} \biggl [(1 - \cosh (\mu L)) 
\frac {2 H^2}{{\mu}^2} + e^{-\mu L} - 1 \pm \nonumber \\
& & \sqrt {((1 - \cosh (\mu L)) 
\frac {2 H^2}{{\mu}^2} + e^{-\mu L} - 1)^2 + N^2 \sinh (\mu L) (\frac {2 H^2}
{{\mu}^2} + 2 e^{-\mu L})} \biggr ]\label {deltrhol}
\eea
In \rf{deltrhol} the solution with plus sign gives $\Delta {\rho}'_L 
\approx -2$ which deviates from our first assumption $|\Delta {\rho}'_L| < 1$ 
and leads to a negative tension on the visible brane like static RS model. 
The solution with negative sign is:
\be
\Delta {\rho}'_L \approx 2e^{-\mu L} \label {deltrholp}
\ee
and both branes have positive tension close to $\lambrs$.\\
When the matter densities on the branes and in the bulk are 
negligible~\cite{houribr}, $\dot {a}^2_0 / a^2_0 = \dot {a}^2_L / a^2_L = 
H^2$ and:
\be
\frac {{\mathcal C}(t)}{a_0^2} \equiv {\mathcal C}' = - \frac {2 H^2}
{{\mu}^2} \label {cp}
\ee
It is easy to see that ${\mathcal D}$ and $a^2 (t, y)$ have the same roots. 
Models with a horizon i.e. a point $y_h$ such that $a^2 (t, y_h) = 0$ 
are pathological (because no particles/brane behind it is observable). The 
condition to have no real root i.e no horizon in the bulk is:
\be
-2 + \frac {{\mathcal C}'^2}{2} \leqslant \Delta {\rho}'_0 + {\mathcal C}' 
\leqslant -\frac {{\mathcal C}'^2}{2} \label {dpos}
\ee
For massive particles, the denominator of the integrand in \rf{tstop} can 
have two roots:
\be
z_{\pm} = \frac {{\mathcal C}' - {\theta}^2 \pm \sqrt {({\mathcal C}' - 
{\theta}^2)^2 + (2 + \Delta {\rho}'_0 + {\mathcal C}')(\Delta {\rho}'_0 + 
{\mathcal C}')}}{\Delta {\rho}'_0 + {\mathcal C}'} \label {zroots}
\ee
The model is consistent only if ${\mathcal D} (z - {\mathcal D} /{\theta}^2) 
> 0$ in the range of integration. Therefore:
\be
z_+ \leqslant z_0 = e^{\mu L} \leqslant z_-. \label {zrange}
\ee
The matter on the brane is confined only if $z_+ > 1$ and $z_+ 
\rightarrow z_0$ when $u^4 \rightarrow 0$. To first order in $e^{-\mu L}$ and 
$N^2$ this leads to the following relation between parameters of the model: 
\be
-{\mathcal C}' = \Delta {\rho}'_0 + \frac {2}{z_0} (N^2 + \Delta {\rho}'_0) 
\label {h2cond}
\ee
This condition is not an addition to the model described in ~\cite{houribr}. 
It is in fact the result of solutions \rf{deltrho0p} and \rf{deltrholp} 
under the approximations considered here. It is not evident whether such a 
constraint appear in the full theory.\\
For $\mu L \gtrsim 5$ the right hand side of \rf {h2cond} is positive. 
Therefore $-{\mathcal C}' \propto H^2$ {\bf can not be zero}. This relation 
between a 
small but non-zero value of the Hubble Constant or equivalently Cosmological 
Constant on the visible brane and the smallness 
of $N^2$ and $\mu$ which is related to the strength of the induced 
gravitational coupling on the brane, confirms the same observations in 
~\cite{houribr} for an  analytical solution of 2-brane models with some approximations and in ~\cite{tye} for the 
exact solution of some special models.\\
Finally the propagation time in the bulk is given by:
\bea
\Delta t_{propag} & = & \frac {4}{\mu (8 |\Delta 
{\rho}'_0 + {\mathcal C}'|)^{\frac {1}{4}}} {\mathcal F} (\alpha, Q). 
\label {tprog2} \\
\alpha & = & 2 \arctan \sqrt \frac {q (z_- - z_0)}{p (z_0 - z_+)} \label 
{alphdef} \\
Q & = & \frac {1}{2} \sqrt {2 + \frac {2}{pq}\biggl [
\frac {{\mathcal C}'({\mathcal C}' - {\theta}^2) + 4 (2 + \Delta {\rho}'_0 + 
{\mathcal C}')(\Delta {\rho}'_0 + {\mathcal C}')}{(\Delta {\rho}'_0 + 
{\mathcal C}')^2} \biggr ]}. \label {qdef} \\
p^2 & \equiv & \biggl ( \frac {{\mathcal C}'}{\Delta {\rho}'_0 + 
{\mathcal C}'} - z_-\biggr )^2 + r^2, \quad \quad q^2 \equiv \biggl ( \frac 
{{\mathcal C}'}{\Delta {\rho}'_0 + {\mathcal C}'} - z_+\biggr )^2 + r^2, 
\nonumber \\
r^2 & \equiv & -\biggl ( \frac {{\mathcal C}'}{\Delta {\rho}'_0 + 
{\mathcal C}'} \biggr )^2 - \biggl (\frac {2 + \Delta {\rho}'_0 + 
{\mathcal C}'}{\Delta {\rho}'_0 + {\mathcal C}'} \biggr ). \label {pqr}
\eea
\begin{figure}[t]
\begin{center}
\psfig{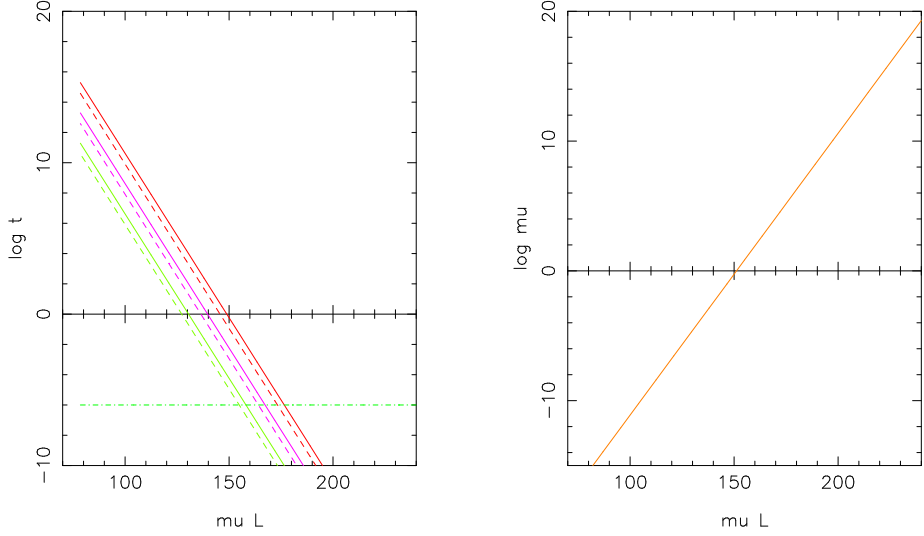}
\caption{Left: Propagation time for relativistic particles in 2-brane model. 
Description of the curves is the same as Fig.\ref{fig:rstime}. 
Right: Parameter $\mu (eV)$ as a function of $\mu L$. It is roughly 
independent of $M_5$ (See the text).\label {fig:gtime}}
\end{center}
\end{figure}
Fig.\ref{fig:gtime} shows the propagation time for models which satisfy 
simultaneously \rf{deltrho0p}, \rf{deltrholp} and \rf{h2cond}. In 
equation \rf{h2cond} up to first order, ${\mathcal C}'$ depends only on 
$\mu L$ and thus in \rf{cp} the value of $\mu$ is independent of $M_5$. 
Only models with large $\mu L \gtrsim 150$ are not ruled out. This is due 
to \rf{dpos} and smallness of the observed Hubble Constant $H$. 
The same conditions make $\mu > 1 eV$ which is much higher than the 
$\mu$ for the fine-tuned RS model. 
With $\mu \lesssim m_{radion}$~\cite {radion} ~\cite {radionpert} these 
class of brane models are 
also consistent with constraint on the fifth-force measurements~\cite {fifth}. 
From \rf {zeromode} the corresponding life-time for SM fields with $m_5 > 0$ 
is shorter than $10^{-16} sec$, much shorter than propagation time in the 
atmosphere and also much shorter than propagation time in the bulk. This 
justifies the classical treatment of propagation. 

We have also applied the formalism described here to the fine 
tuned model of ~\cite{houribr}. In this model the equation of state on the 
hidden 
brane is fine-tuned to neutralize its effect on the visible brane. This 
results to a unique definition of $G$ and the only free parameter in the 
model is $M_5$. Although this model 
has been obtained just by phenomenological arguments, a field theory model 
suggested by Arkani-Hamed \etal~\cite{nima0} to solve the Cosmological 
Constant problem has the same form for $G$ if $\mu L \sim \ktwo \phi (0)$ 
where $\phi (0)$ is the vev of radion on the visible brane (The 
Arkani-Hamed \etal  model has only one brane but it includes a horizon in 
the bulk which limits the accessible size of the extra-dimension and makes 
it similar to a two brane model).

\begin{figure}[t]
\begin{center}
\psfig{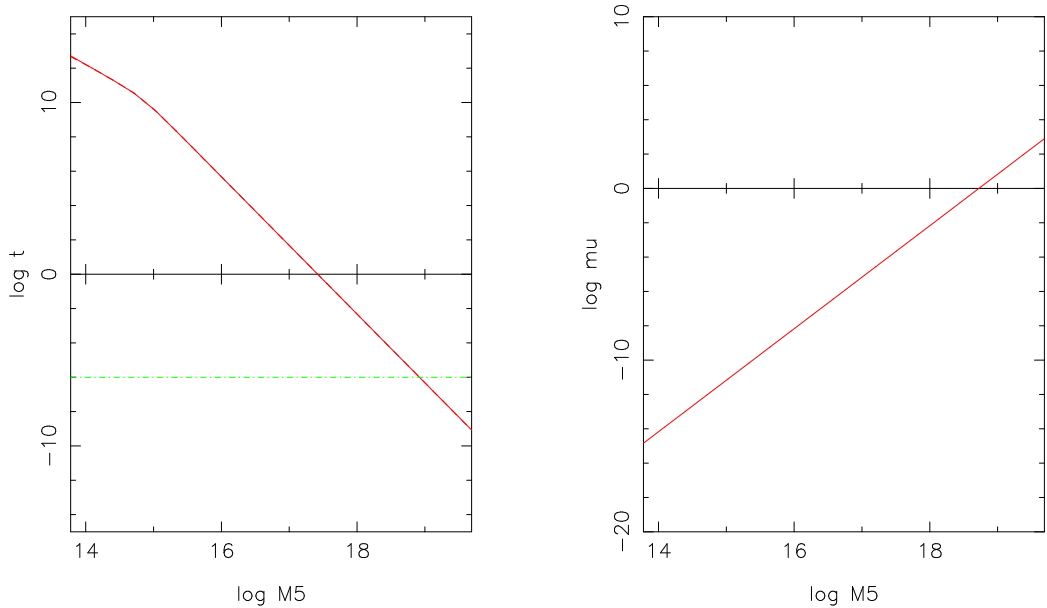}
\caption{Left: Propagation time for relativistic particles in the fine-tuned 
2-brane model of ~\cite{houribr}. Right: Parameter $\mu$ as a function of 
$M_5$.\label {fig:ggtime}}
\end{center}
\end{figure}

\begin{figure}[t]
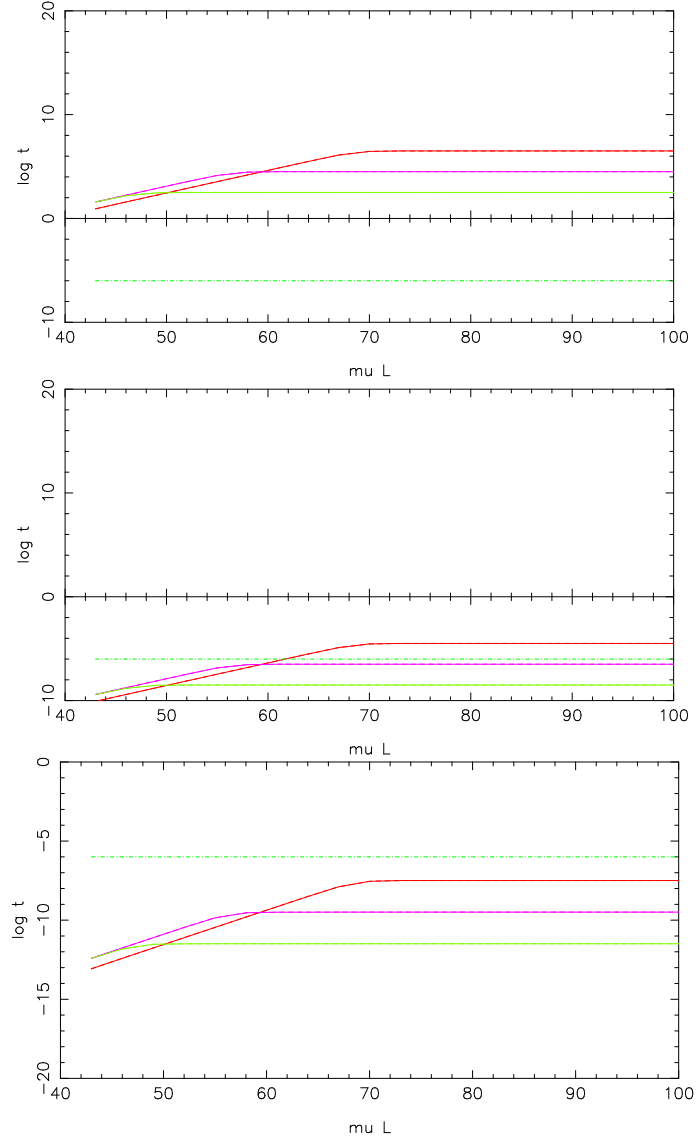

\begin{center}
\epsfig{figure=propagmu-7.eps,width=5cm,angle=-90}
\epsfig{figure=propagmu4.eps,width=5cm,angle=-90}
\epsfig{figure=propagmu7.eps,width=5cm,angle=-90}
\caption{Propagation time for relativistic particles in 2-brane models 
with $\mu$ as a free parameter. Top left: $\mu = 10^{-7} eV$; Top right: 
$\mu = 10^{4} eV$; Bottom: $\mu = 10^7 eV$. Description of the curves is the same as Fig.\ref{fig:rstime}. \label {fig:gmu7time}}
\end{center}
\end{figure}

Fig.\ref{fig:ggtime} shows the propagation time and $\mu$ as a function of 
$M_5$. Unfortunately despite physical interest of this model it is 
only compatible with very high $M_5 \gtrsim 10^{19} eV$ unless the decay to 
the bulk is prevented up to such high energies. The corresponding 
$\mu$ however is consistent with present constraint on the Fifth 
force~\cite {fifth} and the life-time of zero-modes of massive 5-dim. fields 
for such high $M_5$ is enough short to permit particles to decay to the bulk 
during their propagation in the atmosphere. The value for $M_5$ is some 6 
orders of magnitude larger than $10 TeV$, presumably the Electroweak 
interaction scale and it would be a matter of speculation to consider this 
model as having no hierarchy problem. Another problem in testing these models 
with UHECRs is that as the natural scale of gravity $M_5$ is very high, it is 
possible that the symmetry breaking scale which is necessary for the 
localization of fermions (and indirectly gauge bosons) is also much higher 
than CM energy of UHECRs interaction. In this case only very weakly 
interacting particles like gravitons can decay to the bulk. As the total 
production cross-section for them can be tiny, the number of observed UHECRs 
event can be not enough to constrain such models.
We have also tested the general 2-brane models without taking into account 
\rf{h2cond}. Roughly speaking, it is equivalent to having a comparable 
matter density and tension on the hidden brane. The value of $\mu$ becomes a 
free parameter. The result is shown in Fig.\ref{fig:gmu7time} for 3 
different values of $\mu$. Models with $\mu \gtrsim 10^4 eV$ and 
$\mu L \lesssim 70$ are compatible with the present observation of UHECRs. 
The lower limit for $\mu$ from this test is higher than the constraint 
obtained from Fifth force experiments~\cite {fifth}.

\subsection {One-Brane Models}
\begin{figure}[t]
\begin{center}
\psfig{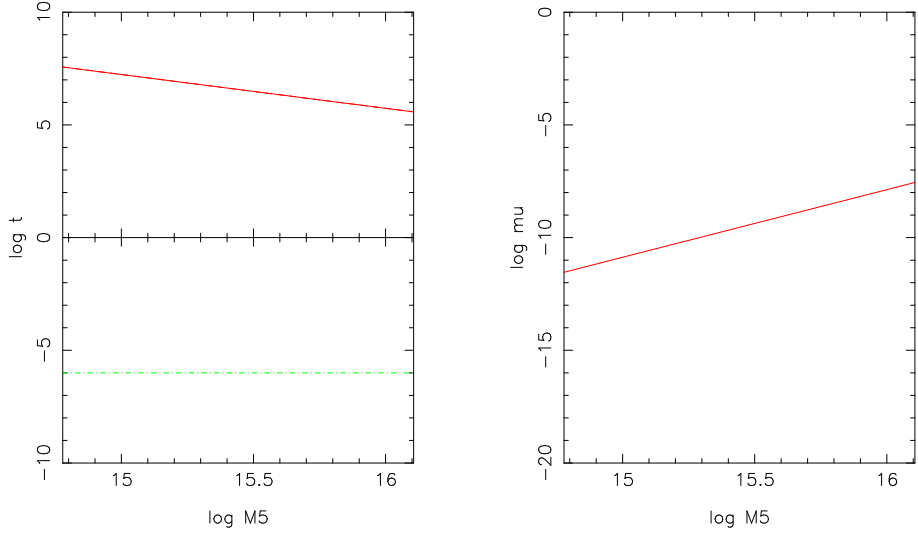}
\caption{Left: Propagation time 
for one-brane models of ~\cite{binetruy99}. Right: Parameter $\mu$ as a 
function of $\mu L$. \label {fig:onebint}}
\end{center}
\end{figure}

The solution of Einstein equations for symmetric one-brane models is the 
same as two-brane ones~\cite{binetruy99}. Due to existence of only one 
boundary however the bulk and brane tensions are not related. In addition,  
the cosmological evolution on the brane:
\be
\frac {\dot {a}_0^2}{a_0^2} = \frac {\ktwo}{6} {\rho}_B + \frac {\kfour}{36} 
 ({\rho}_b + {\rho}_m (t))^2 + \frac {C (t)}{a_0^4}. \label {evolone}
\ee
includes an arbitrary function $C (t)$ which is related to the bulk tension 
and matter~\cite{kanti99}. Here we test two popular models studied in 
~\cite{binetruy99} and ~\cite{kanti99}.\\
In the first model~\cite{binetruy99} $C (t) = 0$ and the brane tension is 
fine-tuned to cancel the effect of quadratic term at late times:
\be
\frac {\ktwo}{6} {\rho}_B + \frac {\kfour}{36} {\rho}_b^2 = 0 \label {biecon}
\ee
and:
\be
8\pi G = \frac {\kfour {\rho}_b}{6} \label {gone}
\ee
The only free parameter in the model is $M_5$. In the second 
model~\cite{kanti99} $C (t)$ (or equivalently $T^5_5$ 
component of the energy-momentum tensor) is adjusted such that the 
conventional evolution equation be obtained. At late times when the brane 
tension is much larger than time dependent matter terms these two models are 
roughly the same.\\
Equations \rf{biecon} and \rf{gone} determine ${\rho}_b$ and $\mu$. It is 
easy to see that ${\rho}_b = \lambrs$. The definition of ${\mathcal C}'$ and 
roots are the same with $\Delta {\rho}_0 = \Delta {\rho}_b = 0$ and 
$z_0 = 1$. 
Fig.\ref {fig:onebint} shows the propagation time for these models.\\
\subsubsection {Effect of ${\theta}^i \neq 0$}
When ${\theta}^i \neq 0$ equations \rf{u0eqpa} and \rf{u4eqpa} can be 
written as:
\bea
\frac {du^0}{d\tau} + \frac {2u^0}{n}\frac {dn}{d\tau} + \frac {\dot a 
{\theta}^i {\theta}^j {\delta}_{ij}}{n^2 a^3} = 0 & & \label {u0eqpathet}\\
u^4 \frac {du^4}{d\tau} - \frac {a' {\theta}^i {\theta}^j {\delta}_{ij}}
{a^3} u^4 + (u^0)^2 n \frac {dn}{d\tau} = 0 & & \label {u4eqpathet}
\eea
At least formally Eq.\rf{u0eqpathet} can be solved analytically: 
\bea
u^0 & = & \frac {\theta}{n^2} + \frac {{\theta}^i{\theta}^j {\delta}_{ij} {\mathcal G}(t,y)}{n^2} 
\label {u0thet} \\
{\mathcal G} (t,y) & = & \int d\tau \biggl (-\frac {\dot a}{a^3} \biggr ) \label {gdef}
\eea
We don't need to solve \rf {u4eqpathet} directly. Knowing $u^0$ and $u^i$ 
we can use the definition of velocity vector to determine $u^4$:
\be
n^2 (u^0)^2 - a^2 u^i u^j {\delta}_{ij} - (u^4)^2 = \varepsilon \label {vel}
\ee
The definition of $\varepsilon$ is the same as in \rf {dydt}. After 
elimination of $d\tau$ we obtain the formal description of the equation of motion 
in the bulk:
\be
\frac {dy}{dt} = \frac {\biggl [\frac {{\theta}^2}{n^2} - \varepsilon + 
{\theta}^i{\theta}^j {\delta}_{ij} 
\biggl (\frac {{\mathcal G}^2 {\theta}^i{\theta}^j {\delta}_{ij}}{n^2} + \frac {2 {\mathcal G} \theta}{n^2} - \frac {1}{a^2} \biggr )\biggr ]^{\frac {1}{2}}}{\frac {\theta}{n^2} + \frac {{\mathcal G}{\theta}^i{\theta}^j {\delta}_{ij}}{n^2}}
 \label {dydtthet}
\ee
We can use \rf {uisol} to determine $d\tau$ (As before for simplicity we 
assume that $Dx^i/d\tau \approx dx^i/d\tau$):
\be
d\tau = \frac {a^2 {\delta}_{ij}{\theta}^j dx^i}{{\theta}^i{\theta}^j {\delta}_{ij}} \label {tausol}
\ee
and:
\be
{\theta}^i{\theta}^j {\delta}_{ij}{\mathcal G} (t(\tau),y(\tau)) = - \int 
\frac {\dot a}{a}{\delta}_{ij}{\theta}^j dx^i \label {gsol}
\ee
In \rf{gsol} $\dot {a} / a$ is independent of $x^i$. The rest of right 
hand side of \rf{gsol} i.e. 
$\int {\delta}_{ij}{\theta}^j dx^i$ is the projection of the particles world 
line on the brane. The value of Hubble constant $\dot {a}(t,y) / a (t,y)
\sim H^2$ (from (\ref{aaa}-\ref{ccc}) and (\ref{dccc}-\ref{dbbb})) and thus 
${\theta}^i{\theta}^j {\delta}_{ij}{\mathcal G}$ is very small 
when the projection distance traversed by the particle is small with 
respect to the Hubble radius. Therefore we presume that conclusions of the 
previous section will not be extremely modified when full propagation is 
considered.

\section {Conclusion}
The calculation in this work is mainly based on two assumptions:
\begin {description}
\item {- } At interaction energy scale of most energetic Cosmic Rays the 
physics is high dimensional either because all symmetry based confinements 
are no longer at work or because there is a large cross-section and/or phase 
space volume which permits the production of massive KK-modes. 
\item {- } In the time scale of the propagation of a particle in the 
extra-dimension, the bulk and the branes are quasi-static.
\end {description}
If these assumptions are valid the time coherence of Ultra High Energy 
Air Showers rules out a large part of the parameter space for a number of 
brane models unless some micro-physics phenomena confine particles to the 
brane at energies much higher than Electroweak scale. This makes a new 
hierarchy inconsistent with the spirit of brane models. 

For most 2-brane models the acceptable range of $\mu$ is 
$\mu \gtrsim 1 eV$ except for original RS model which needs 
$\mu \gtrsim 10^{-2} eV$. This lower limit is much lower than what can be 
obtained from non-observation of KK-mode production in 
accelerators~\cite{ftrs}. However, at present accelerator energies it is 
always arguable that presence of some quantum conservations prevent the 
production of KK-modes. Presence of such conservations at the CM energy of 
UHECR interaction seems much less natural and constraints on $M_5$ more are 
robust. 

The upper limit of $L$ is also a universal value for models with different 
range of $\mu L$: $L \lesssim 10^{-3} eV^{-1} \sim 10^{-8} cm$. It 
is much smaller than the upper limits 
obtained from gravity experiments~\cite{grex}~\cite{fifth}. Again for static 
RS model the upper limit is $L \lesssim 10^5 cm$, much larger than one 
obtained for other models. One brane models with interesting range of $M_5$ 
are ruled out.

This study has an additional interesting conclusion: 
The close relation between a very small but non-zero Cosmological Constant 
and the smallness of the Newton coupling constant (i.e. the hierarchy 
problem). In fact without the fine-tuning of these apparently 
independent physical quantities, the brane models are not consistent.

\end {document}